\documentclass[reqno,prl,showpacs,twocolumn,superscriptaddress]{revtex4-1}
\usepackage{amssymb,latexsym,mathrsfs,footnote}
\usepackage{verbatim}
\usepackage{amsfonts}
\usepackage{amsmath}
\usepackage{color}
\usepackage{graphicx}
\usepackage{array}
\usepackage{multirow}

\newcommand{\beq}{\begin{equation}}
\newcommand{\eeq}{\end{equation}}

\newcommand{\id}{\textrm{d}}

\def\bea{\begin{eqnarray}}
\def\eea{\end{eqnarray}}
\def\ba{\begin{array}}
\def\ea{\end{array}}

\def\la{\langle}
\def\ra{\rangle}

\def\ve{\varepsilon}
\begin{document}

\title{Measurement of second-order response without perturbation}

\author{Laurent Helden}
\affiliation{2. Physikalisches Institut, Universit\"at  Stuttgart, 70550 Stuttgart, Germany}
\author{Urna Basu}
\affiliation{SISSA - International School for Advanced Studies and INFN,  Trieste, Italy}

\author{Matthias Kr\"uger}
\affiliation{4th Institute for Theoretical Physics, Universit\"at
Stuttgart, Germany}
\affiliation{Max Planck Institute for Intelligent Systems, 70569 Stuttgart, Germany}

\author{Clemens Bechinger}
\affiliation{2. Physikalisches Institut, Universit\"at  Stuttgart, 70550 Stuttgart, Germany}
\affiliation{Max Planck Institute for Intelligent Systems, 70569 Stuttgart, Germany}

\begin{abstract}

We study the second order response functions of a colloidal particle being subjected to an anharmonic potential. Contrary to typical response measurements which require an external perturbation, here we experimentally confirm a recently developed approach where the system's susceptibilities up to second order are obtained from the particle's equilibrium trajectory [PCCP {\bf 17}, 6653 (2015)]. The measured susceptibilities are in quantitative agreement with those obtained from the response to an external perturbation.
\end{abstract}


\maketitle
The fluctuation-dissipation theorem (FDT) is a powerful tool in statistical physics, which allows to calculate the response of quantum or classical systems to an external perturbation from its equilibrium fluctuations \cite{Kubo12}. More recently, it has been shown, that this concept can be also applied to nonequilibrium systems  \cite{Harada05, Speck06, Blickle07, Chetrite08, Baiesi09, Prost09, Krueger09}. Because the FDT is limited to the linear response regime which
restricts its validity to small perturbations, higher order response functions are typically derived from non-equilibrium correlation functions \cite{Yamada67,Evans88,Fuchs05}.
During the last decades, several attempts have been made to extend the idea of the FDT to  nonlinear response functions \cite{Kubo54,Lucarini12}. When trying to connect the nonlinear response with equilibrium correlation functions, however, a fundamental difference compared to the FDT appears: While the evaluation of linear response functions only requires knowledge of the perturbation, the evaluation of second (or higher) order response necessitates additional information about the interactions and dynamics of the system \cite{Basu2015}. So far, however, this concept has not been demonstrated experimentally and it is not clear, whether with this concept nonlinear response functions can be measured with high accuracy in equilibrium experiments.

In this Letter we demonstrate that the second order response of a micron-sized colloidal particle in an anharmonic potential can be obtained
solely from its experimentally measured equilibrium fluctuations. Compared to conventional perturbation measurements, this approach has the advantage, that it allows to predict the second-order response to arbitrary perturbation protocols from a single experiment.

{\it Theory} -- The overall goal of response theory as used here, is to predict the reaction, i.e. the susceptibility, of an observable $O$ to a perturbation from the system's equilibrium correlation functions. $O$ is typically a function of  phase space variables $\mathbb{X}_t$ and will depend on time $t$. For small perturbation amplitudes, $O$ can be expanded in orders of $\varepsilon$  which quantifies the perturbation magnitude relative to the equilibrium forces acting in the system  \cite{Basu2015},
\bea
\label{eq:2nd}
\lefteqn{\la O(\mathbb{X}_t) \ra_\ve - \la O(\mathbb{X}_t) \ra_0 =}  \nonumber  \\
& & \varepsilon \la {\cal S} O(\mathbb{X}_t) \ra_0-\varepsilon^2 \la {\cal S} {\cal D} O(\mathbb{X}_t) \ra_0 + \mathcal{O}(\varepsilon^3). 
\eea     
Here, $\la \cdot \ra_0$ and $\la \cdot \ra_\ve$ correspond to the averages of the equilibrium and perturbed system, respectively. The linear response, i.e., the first term in the second line of Eq.~\eqref{eq:2nd}, depends on the so-called {\it excess entropy} ${\cal S}$ , which must be evaluated along the trajectories. To be more precise, the linear response is the antisymmetric part of the action (i.e. the probability to find a certain trajectory) assigned to perturbed paths which can be also related to the work done by the perturbing force along the corresponding trajectory. It can be immediately evaluated when the perturbation imposed on the system is known. Accordingly, when limiting to first order, equation~\eqref{eq:2nd} is nothing but the formulation of the well-known Onsager principle: a system responds to an external perturbation in the same manner as to equilibrium fluctuations.

The nonlinear response, as given by the second term in the lower line of Eq.~\eqref{eq:2nd}, involves the so-called {\it dynamical activity} $\cal{D}$ \cite{Baiesi09}. In contrast to $\mathcal{S}$, $\cal{D}$ corresponds to that contribution of the action assigned to a perturbed path which is symmetric under time reversal. In addition to the perturbation, it depends on the details of the system's dynamics (as specified below). Notably, Eq.~\eqref{eq:2nd} implies that the Onsager principle can be extended to second order. Despite its implications and practical use, so far it has not yet been demonstrated, whether the path function $\cal{D}$ is experimentally accessible and whether equilibrium fluctuations are sufficiently strong and frequent, to explore the nonlinear response regime.

\begin{figure}[t]
\centering
\includegraphics[width=8.8 cm]{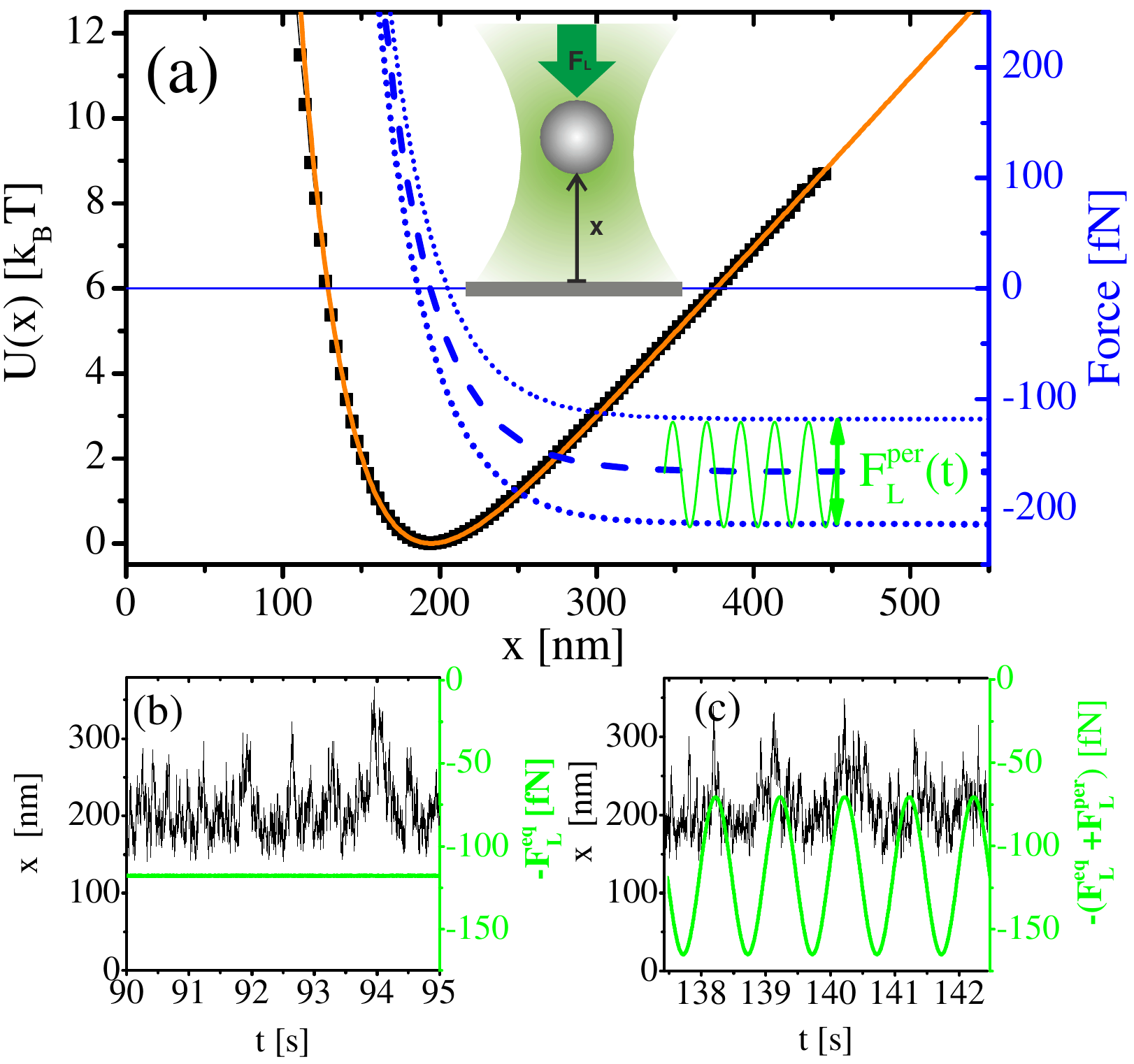} 
 \caption{(a) Measured particle-wall interaction potential (solid squares, left scale) and fit to Eq.~\ref{eq:pot}  with $U_0=550 k_BT$, $\lambda_D=32\pm1\mathrm{nm}$ and $F_G+F_L^\text{eq}=166\pm4$~fN (solid line (orange)). The dashed line (blue) shows the corresponding force $ -\nabla U(x)$ (right scale). The dotted lines (blue) show the force in case of driving,  at maximum  and minimum modulation strength $(h_t=\pm1)$. The inset sketches the particle near the wall within the optical tweezer. (b) Section of trajectory (line (black), left scale)  corresponding to the potential in (a) at constant applied light force (solid line (green/grey), right scale). (c) Same as (b) but for modulated light force with amplitude $\varepsilon=0.403$ and $T_p=1\mathrm{s}$ period. }
\label{fig:pot}
 \end{figure}

{\it Experimental Setup} -- In order to address these questions and to demonstrate the validity of Eq.~\eqref{eq:2nd} up to second order, we studied the motion of a colloidal particle with radius $r=1.32~ \mathrm{\mu m}$ dispersed in aqueous solution near a flat glass wall (Fig.~\ref{fig:pot}a). For the coordinate of interest $x$ perpendicular to the wall, the particle was confined by an anharmonic potential $U(x)$. The potential results from the electrostatic repulsion between the negatively charged surfaces of the particle and the wall, the gravitational force $F_G$ acting on the particle and a constant light force $F_L^\text{eq}$ \cite{Walz1992}. The latter was created by an optical tweezer, i.e. a weakly focused laser beam which is incident from the top (inset Fig.~\ref{fig:pot}a). The total potential is

\beq
 U(x) = U_0 \exp{(-x/\lambda_D)} +(F_G + F_L^\text{eq}) x 
\label{eq:pot}
\eeq
where the strength and range of the electrostatic interaction are denoted by $U_0$ and $\lambda_D$, the latter corresponding to the Debye screening length \cite{Gru01-1}.

The particle trajectory $x_t$ in such an asymmetric potential was measured using the method of total internal reflection microscopy (TIRM) with a temporal and spatial resolution of $1$~ms and $2~\mathrm{nm}$, respectively (see Fig.~\ref{fig:pot}(b)). The lateral particle motion was strongly suppressed by the optical tweezer \cite{Walz1992}. For details regarding TIRM we refer to the literature~\cite{Blick2006,Prieve1999,Walz1997}. From the particle's trajectory one obtains the probability distribution $P(x) \propto e^{-\beta U(x)}$ which finally yields the potential $U(x)$. Here $\beta=(k_B T)^{-1}$ with $T$ the temperature and $k_B$ the Boltzmann constant. In our experiments $T$ was kept constant at $T= 299.7\pm 0.1$~K. The solid squares in Fig.~\ref{fig:pot}(a) represent the measured $U(x)$ which indeed is well described by Eq.~\ref{eq:pot} (solid line). Here, $F_G=Vg\Delta\rho=48$~fN, was obtained from the density difference  $\Delta \rho$ between the particle and the solvent and the particle volume $V$. From the fit we also obtained $U_0$, $\lambda_D$ and $F_L^\text{eq}$ whose values are given in the caption of Fig.~\ref{fig:pot}.

To characterize the colloidal dynamics which is crucial for the dynamical activity $\cal{D}$, we have measured the particle's diffusion coefficient $D$ perpendicular to the wall which can be directly obtained from $x_t$ ~\cite{Oetama2005}.  Due to hydrodynamic interactions, $D(x)$ is known to fall below the corresponding (bulk) Stokes-Einstein value $D_0$ at smaller particle-wall distances \cite{Brenner1961,Bevan2000}
\beq \label{eq:Diff}
D(x) \cong D_0 \frac{6x^2 + 2rx}{6x^2 + 9rx+2r^2}.
\eeq 
For $D_0=0.196 \mathrm{\mu m^2/s}$ corresponding to a colloidal particle with $r=1.32~ \mathrm{\mu m}$, this distance-dependence is in good agreement with our data.

To compare response functions obtained from equilibrium trajectories (Eq.~\eqref{eq:2nd}) with those resulting in the presence of an external perturbation, an additional time-dependent light force $F_L^\text{per}(t)$ has been applied to the particle. Since the light force acting on the particle is proportional to the intensity of the optical tweezer, experimentally, this was achieved by modulating the laser intensity by means of an acousto-optical modulator. It is controlled by a feedback-loop which guarantees an accuracy and long-time stability of the transmitted intensity better than $10^{-3}$. Accordingly, the total light force acting on the particle is $F_L^{tot}(t)= F_L^\text{eq}+F_L^\text{per}(t)$. To demonstrate the effect of the perturbing force on the particle motion, in Fig.~\ref{fig:pot}(b,c) we have plotted particle trajectories with a constant and a modulated light force. In the latter case, a clear correlation between $F_L^\text{per}(t)$ and the particle position is observed \footnote{ \label{SignOfForce} Note that $F_L^\text{per}$ has the same sign convention as $F_L^\text{eq}$ and $F_G$, i.e., the force is pointing downwards in negative $x$-direction.}. On average, the perturbation force leads to a shift of the particles position of only a few nm towards larger distances.

{\it Data analysis} -- 
The calculation of response functions to an external perturbation requires the exact knowledge of the perturbation protocol. In our experiments we have chosen a sinusoidal function $h_t = \sin(\omega_p t)$ with period $T_p = \frac{2\pi}{\omega_p}$. This leads to a time-dependent perturbation force $F_L^\text{per}(t)=\varepsilon F_L^\text{eq} h_t$, where $\varepsilon$ specifies the amplitude of the perturbation in units of the equilibrium light force $F_L^\text{eq}$.  Then, the excess entropy $\cal S$ in Eq.~\eqref{eq:2nd} is given by \cite{Basu2015}
\bea\label{eq:S}
 \mathcal{S}(t) = -\beta F_L^\text{eq} \left[h_t x_t -h_{t_0}x_{t_0} - \int_{t_0}^t\id s\, \dot{h}_s x_s \right],
\eea
where $t_0$ denotes the starting time of the perturbation protocol. As already mentioned, $\cal S$ and hence the linear response, only requires knowledge of the perturbing force. This is in contrast to the dynamical activity $\cal{D}$ which requires additional information about the system. In the case considered here, this is the potential $U(x)$ and the distance-dependent diffusion coefficient $D(x)$. Then $\cal{D}$ is given by \cite{Basu2015}

\bea
\hspace{-5mm} 
\mathcal {D}(t)&=&-\frac{\beta F_L^\text{eq}} 2\int_{t_0}^t\id s\, h_s\, [-\beta D(x_s) \nabla U (x_s)+ \nabla D(x_s)]. \label{eq:SD0}
\eea

When considering the particle's position as the observable $O(\mathbb{X}_t)\equiv x_t$, we define the linear and second order susceptibility $\chi_1$ and $\chi_2$ as 
\bea
\la x_t \ra_\ve -\la x_t \ra_0 = \varepsilon \chi_1(t)   + \varepsilon^2 \chi_2(t) + \mathcal{O}(\varepsilon^3) \label{eq:Taylor}
\eea
where $\chi_1$ and $\chi_2$ have the dimension of a length. 

In thermal equilibrium, we identify by comparison with Eq.~\eqref{eq:2nd}
\bea
\chi_1^{\rm eq}& = & \la {\cal S} x_t \ra_0 \hspace{1cm}  \mathrm{and} \label{eq:Chi1}\\
\chi_2^{\rm eq}& = & -\la {\cal S} {\cal D} x_t \ra_0 . \label{eq:Chi2}
\eea
Note that  Eq.~\eqref{eq:Chi2} requires the evaluation of three-time correlation functions which require long sampling times. To achieve well-defined averages of $\chi_2^{\rm eq}$, in our experiments we have analyzed trajectories of about 10 hours duration. 

For perturbed trajectories we separate even and odd powers of
$\varepsilon$ by use of the following forms, which become exact for
sufficiently small $\varepsilon$
\bea
\chi_1^{\text{per}} &:=& \frac 1{2\varepsilon} [ \la x_t \ra_\ve -\la
x_t\ra_{-\ve} ] \label{eq:chi_per1}, \\
\chi_2^{\text{per}} &:=& \frac 1{\varepsilon^2}\left[\frac{\la x_t
\ra_{\varepsilon} + \la x_t \ra_{-\varepsilon}}2 - \la x\ra_0\right]
\label{eq:chi_per2}.
\eea
Here $\la \cdot \ra_{-\ve}$ corresponds to trajectories for opposite sign
of $\ve$. To reduce statistical errors and to directly compare susceptibilities obtained from equilibrium and perturbed data, in the following we have analyzed particle trajectories over up to $10^5$ cycles of the external perturbation protocol.

\begin{figure}[t]
 \centering
  \includegraphics[width=8.8cm]{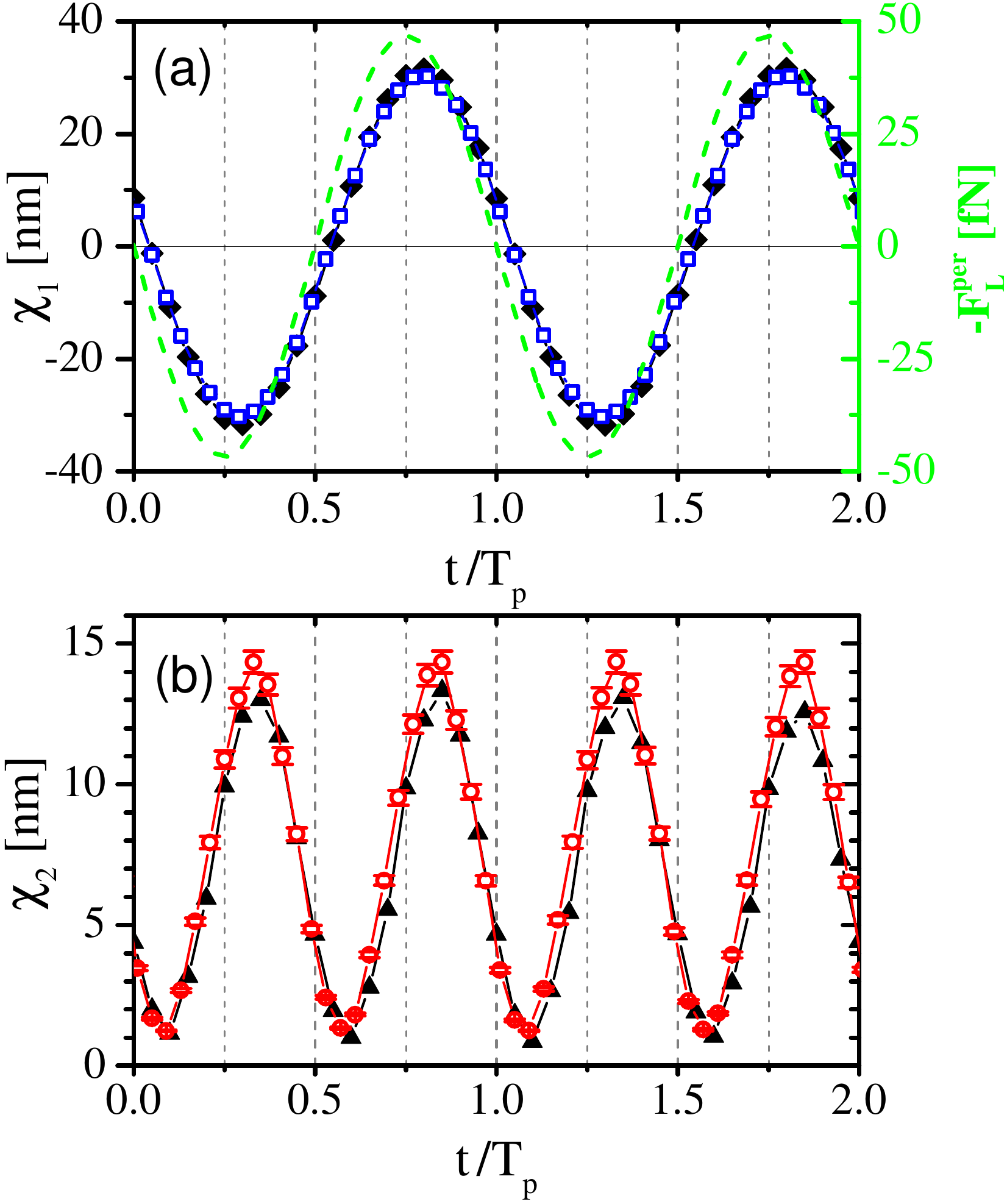}
 \caption{(a) First order susceptibilities $\chi_1^\text{eq}$ obtained in thermal equilibrium (full diamonds, (black)) and $\chi_1^\text{per}$ from externally perturbing the system with $\varepsilon=0.403$ and $T_p=1s$ (open squares (blue)). The perturbation force $-F_L^\text{per}(t)$ is added as dashed line (green).  (b) Corresponding second order susceptibilities $\chi_2^\text{eq}$ plotted as full triangles (black) and $\chi_2^\text{per}$ as open circles (red). Error bars reflect the experimental uncertainties in $\varepsilon, \lambda_D$ and $F_L^\text{eq}$.}
  \label{fig:T1A10}
\end{figure}

{\it Results} -- 
Figure~\ref{fig:T1A10}(a) compares the experimentally determined linear response $\chi_1^\text{eq}$ and $\chi_1^\text{per}$ of a colloidal particle to a periodic perturbation with $\ve=0.403$ and $T_p=1$~s. 
As expected, the response is identical to that of an overdamped harmonic oscillator, i.e., $\chi_1(t)$ is a monochromatic sinusoidal function of $\omega_p t$ with zero mean and a phase shift relative to the driving force \cite{overd}. The phase shift obtained from Fig.~\ref{fig:T1A10}(a) is $-0.044~T_p$. It results from the particle's finite relaxation time $\tau= 46$~ms, as obtained from the decay of the particle's positional autocorrelation function. For the evaluation of $\chi_1^\text{eq}(t)$ (Eq.~\eqref{eq:Chi1}), the lower limit of the integral in Eq.~\eqref{eq:S} was set to $t_0=-T_p$ where $T_ p\gg \tau$. This ensures, that transient effects due to $\tau$ have decayed for $t>0$. The expected agreement between $\chi_1^\text{per}$ and $\chi_1^\text{eq}$ is a direct experimental confirmation of the FDT. 

Fig.~\ref{fig:T1A10}(b) shows the corresponding results for the second order response $\chi_2$ (Eqs.~\eqref{eq:Chi2},~\eqref{eq:chi_per2}) where $\mathcal{D}$ was also evaluated with the lower integration boundary $t_0=-T_p$. Clearly, $\chi_2$ contains  $2 \omega_p$ frequency components, as this is characteristic for second order response (second harmonic generation). It should be mentioned, that the second order response in our system is not only a result of the anharmonic shape of $U(x)$ but also of the distance dependent diffusion coefficient (cf. Eq.~\eqref{eq:SD0}). 
Similar to the linear response, we find in Fig. \ref{fig:T1A10}(b),
that $\chi_2^\text{eq}$ (full triangles) and $\chi_2^\text{per}$ (open circles) agree well and thus experimentally confirm that second order response can be obtained solely from the analysis of equilibrium data. 
 
To substantiate this concept, we compare the response for different driving strength $\ve$ and frequency $\omega_p$. This is most conveniently done by expanding the mean particle position for a given driving frequency $\omega_p=\frac{2\pi}{T_p}$ in a Fourier series up to second order,
\begin{align}\label{eq:F}
\la x_t \ra_\ve - \la x_t \ra_0 = A + B \sin (\omega_p t + \phi_1) + C \sin (2\omega_p t + \phi_2) 
\end{align}
with a time independent response $A$ and the phases $\phi_1$ and $\phi_2$. From symmetry arguments it follows, that the Fourier coefficients $A$ and $C$ are even in the order of $\ve$ whereas $B$ is of odd order. Accordingly, $A/\ve^2$ and $C/\ve^2$ correspond to the time average and oscillation amplitude of $\chi_2$, respectively, while $B/\ve$ is the oscillation amplitude of $\chi_1$ (cf. Fig.~\ref{fig:T1A10}). 

 \begin{figure}[t]
 \centering
 \includegraphics[width=8.8 cm]{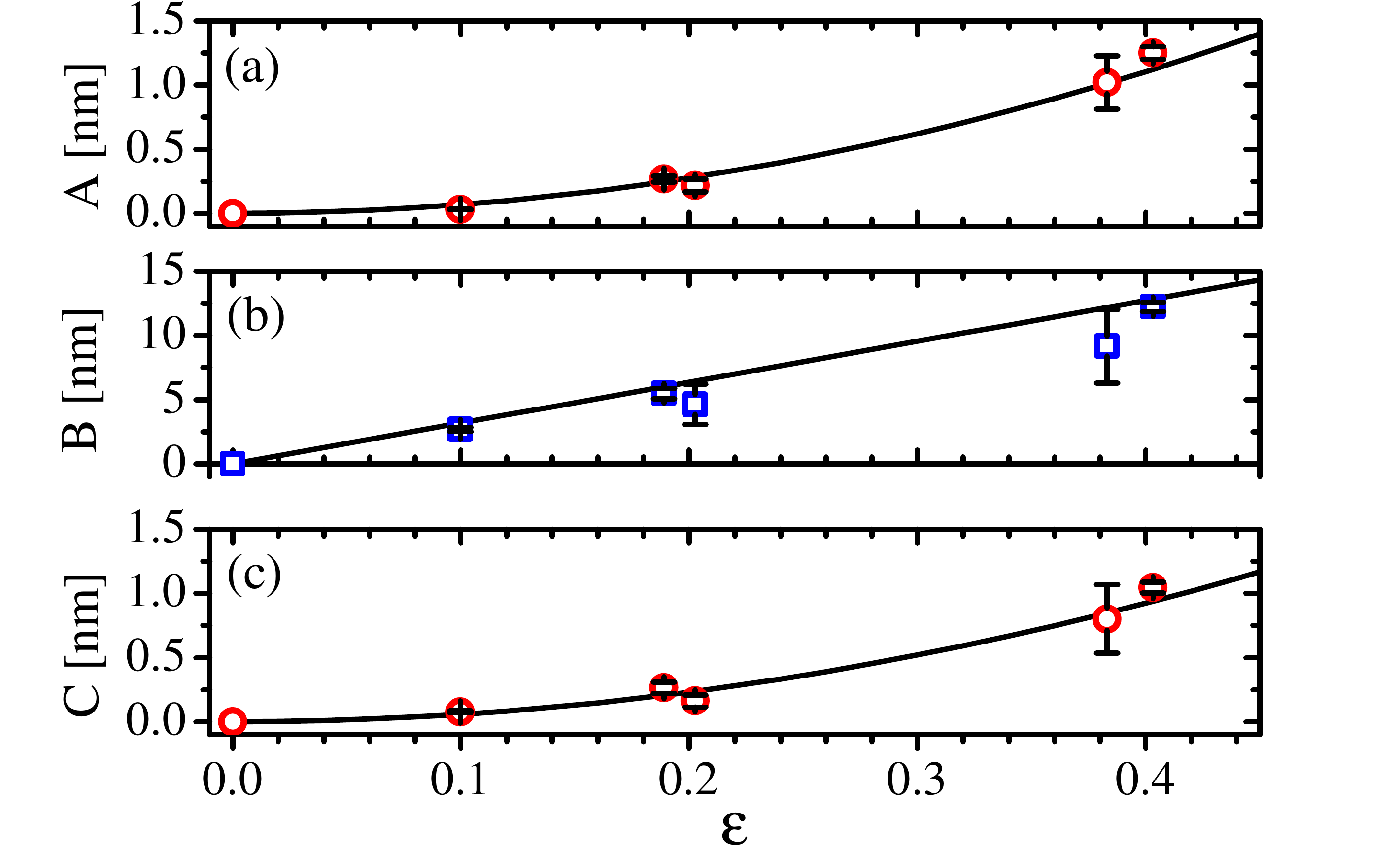}
 
 \caption{Fourier coefficients obtained from equilibrium experiments (lines) and in presence of an external perturbation (open symbols) as a function of the driving strength $\varepsilon$ and for driving period $T_p=1$~s ($\omega_p=2\pi s^{-1}$). The error bars of the perturbed, i.e. non-equilibrium data, correspond to small variations in the Debye screening length between individual measurements.}
 \label{fig:fourier_eps}
\end{figure}
Fig.~\ref{fig:fourier_eps} shows the Fourier coefficients obtained from equilibrium measurements (lines) and in the presence of an external perturbation (open symbols) as a function of the perturbation strength $\ve$ for a modulation time $T_p=1$~s. For the equilibrium data, the curvatures of the parabola in Figs.~\ref{fig:fourier_eps}(a,c) correspond to $A$ and $C$ and are obtained from Eq.~\eqref{eq:Chi2}. The coefficient $B$ varies linearly in $\ve$ with the slope given by Eq.~\eqref{eq:Chi1}. The corresponding parameters as obtained in presence of a perturbation show good agreement with the equilibrium data. The remaining differences are largely due to variations in the Debye screening length which slightly varied between individual measurements. From our experimental non-equilibrium data, we also determined the phases as defined in Eq.~\eqref{eq:F} to $\phi_1=-(0.544\pm0.0002)2\pi$ and $\phi_2=-(0.42\pm0.016)2\pi$.

\begin{figure}[t]
	\centering
	\includegraphics[width=8.8 cm]{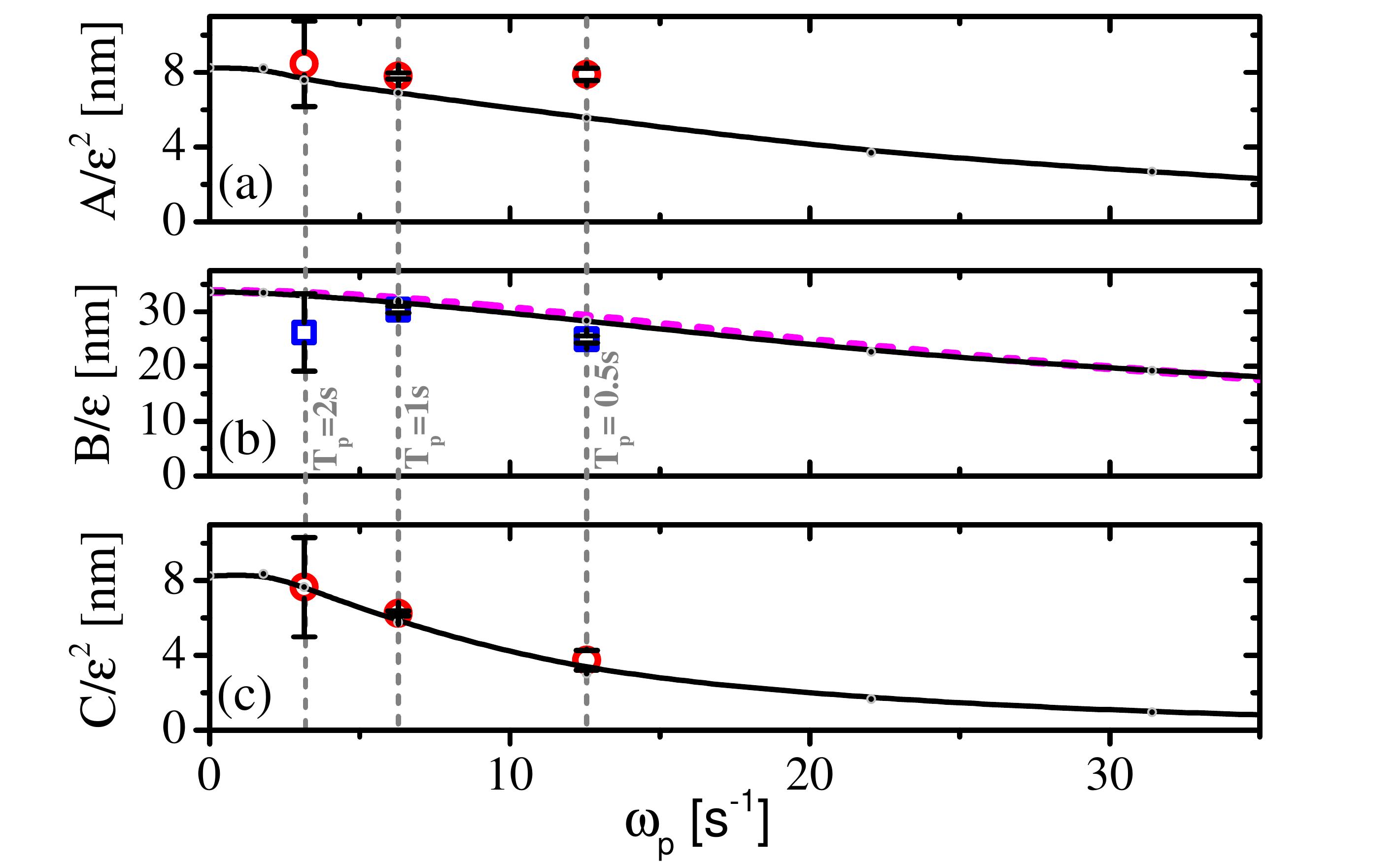}
	\caption{Fourier coefficients as a function of the frequency $\omega_p$ of the external perturbation as obtained from equilibrium data (solid lines) and in presence of a perturbation protocol (open symbols). The dashed line (magenta) in (b) shows the linear response  of an overdamped  oscillator and is in excellent agreement with the corresponding Fourier coefficient characterizing the first-order response.}
	\label{fig:fourier_T} 
\end{figure}

Owing to the asymmetry of the potential $U(x)$ (see Fig.~\ref{fig:pot}) the center of the particle probability distribution is slightly displaced to the right of the potential minimum. As a consequence $A>0$ (cf. Eq.~\eqref{eq:F}). Since $A\approx C$ (for $\omega_p \rightarrow 0$, $A = C$ as seen in Fig.~\ref{fig:fourier_T}), this explains why the minima of $\chi_2$ in Fig.~\ref{fig:T1A10}~(b) are close to zero (cf. Eq.~\eqref{eq:F}). 

 Finally, we also studied the frequency dependence of the second order response functions which is shown in Fig.~\ref{fig:fourier_T}, where the rescaled Fourier coefficients $A/\ve^2$, $B/\ve$ and $C/\ve^2$ are plotted as a function of $\omega_p$. The open symbols correspond to measurements with a time period of the perturbation protocol of $T_p=0.5$, $1$ and $2$~s. For comparison, we also show the data obtained in equilibrium, i.e. without perturbation (solid lines). It should be emphasized, that in this case, the entire frequency dependence is obtained from a single experiment (cf. Eq.~\eqref{eq:SD0}). Apart from the higher accuracy of measurements in thermal equilibrium, this is of considerable advantage when predicting the second order response to an arbitrary perturbation protocol. Again we find an overall good agreement between equilibrium and non-equilibrium data, which confirms the validity of our approach. The values for $\omega\to0$ are in good agreement with the corresponding results obtained from the quasi-static distribution $P(x,t) \propto e^{-\beta [U(x)+xF_L^\text{per}(t)]}$, yielding $A/\ve^2=C/\ve^2=8.25$~nm and $B/\ve=33.7$~nm. The linear response $B/\ve$ is well described by the response of an overdamped oscillator $B(\omega)\propto 1/\sqrt{1+(\tau\omega)^2}$ with the experimentally determined relaxation time $\tau = 46$~ms (dashed line).

{\it Conclusions and Outlook} -- We have experimentally demonstrated, that the second order response of a colloidal particle which is fluctuating in an asymmetric potential can be measured with high accuracy in thermal equilibrium, i.e. without applying an external perturbation. These data are found to be in good agreement with the response where the system was externally perturbed by an oscillating optical force. A major advantage to extract second-order response functions from thermal equilibrium is, that the entire amplitude- and  frequency-dependence is contained in a single experiment and thus allows to predict the response of a system to arbitrary perturbation protocols. We expect, that this approach should be also applicable to higher order response functions and extendable to quantum systems.

The authors thank Christian Maes for helpful discussions and suggestions. 
UB acknowledges the financial support by the ERC under Starting Grant 279391 EDEQS. MK was supported by Deutsche Forschungsgemeinschaft (DFG) grant No. KR 3844/2-1.


%

\end{document}